\documentclass[aps,superscriptaddress,twocolumn,showpacs]{revtex4}

\usepackage{amsmath}

\usepackage{graphicx}
\usepackage{natbib}

\bibpunct{(}{)}{,}{a}{,}{,}

\begin{document}

\title{Clustering of Ions at Atomic Dimensions in Quantum Plasmas}
\author{P. K. Shukla}
\affiliation{International Centre for Advances Studies in Physical Sciences \& Institute for Theoretical Physics,
Faculty of Physics \&  Astronomy, Ruhr-University Bochum, D-44780 Bochum, Germany}
\affiliation{Department of Mechanical and Aerospace Engineering \& Center for Energy Research,
University of California San Diego, La Jolla, CA 92093, U. S. A.}
\email{profshukla@yahoo.de}
\author{B. Eliasson}
\affiliation{International Centre for Advances Studies in Physical Sciences \& Institute for Theoretical Physics,
Faculty of Physics \&  Astronomy, Ruhr-University Bochum, D-44780 Bochum, Germany}
\email{beliass@yahoo.se}
\date{22 November 2012}
\begin{abstract}
By means of particle simulations of the equations of motion for ions interacting among themselves under the influence of newly discovered
Shukla-Eliasson attractive force (SEAF)  in a dense  quantum plasma, we demonstrate that the SEAF can bring ions closer at atomic dimensions. We present simulation results of the  dynamics of an ensemble of ions in the presence of the SEAF  without and with confining external potentials and collisions between the ions and degenerate electrons. Our particle  simulations reveal  that under the SEAF, ions attract each other, come closer and form ionic clusters in the bath of degenerate electrons that shield the ions. Furthermore, an external confining potential produces robust ion clusters that can have cigar-like and ball-like shapes, which remain stable when the confining potential is removed.  The stability of ion clusters
is discussed. Our results may have applications to solid density plasmas (density exceeding $10^{23}$ per cubic centimeters), where
the electrons will be degenerate and quantum forces due to the electron  recoil effect  caused by the overlapping of electron wave functions
and electron tunneling through the Bohm potential, electron-exchange  and electron-exchange and electron correlations associated with electron-$1/2$ spin effect, and the quantum statistical pressure of the degenerate electrons play a decisive role.
\end{abstract}

\maketitle
During the early thirties, there were several discoveries related to non-Coulombic shielded potential distributions that exhibit the role of collective interactions between electrons and ions in electro-chemistry \citep{Debye23}  (viz. electrolytes and colloidal suspensions), solid state \citep{Thomas27,Fermi27} and gaseous \citep{Langmuir29} plasmas, and between neutrons and protons in elementary particle physics \citep{Yukawa35}. The screened non-Coulombic potentials, which were obtained by using linearized theory based on the assumption that the potential energy between the particles is much smaller than the particle kinetic energy, are now known as the Debye-H\"uckel (DH),  Thomas-Fermi (TF), and Yukawa potentials in the context of electro-chemistry and plasma physics, condensed matter physics,  and nuclear physics, respectively. The DH, TF, and Yukawa potentials describe short-range (of the order of the DH radius, the TF radius, and the Yukawa radius, which are fixed by the size of a shielded cloud) repulsive interactions between two particles that  have the same polarity. The DH theory has also been extended to dusty plasma physics where charged dust particles are shielded by non-degenerate electrons and ions. The DH, TF, and Yukawa interaction potentials, which significantly deviate from the long-range Coulomb interaction potential,  have important applications to the
understanding of phase transitions \citep{Phase,Avinash07,Klumov10} in different areas of physical sciences.

In order for charged particles to form ordered structures under the influence of Coulombic, DH, TF and Yukawa repulsive forces, one must confine the like-charged particles in an external potential, so as to bring them to a minimum energy state. Examples include the Wigner crystals \citep{Wigner34} composed of an ensemble of electrons on the surface of liquid helium, ion crystals in laser cooled Paul \citep{Drewsen98} and Penning (electromagnetic) traps \citep{Tan95}, charged dust particle crystals \citep{Wuerker59},  which were formed when like-charged dust particles were kept together via external confining potentials despite short-range Coulombic or shielded Coulombic repulsive forces between charged dust particles. In fact, both electron and ion crystals, as well as crystals of colloidal suspensions  and oil droplets have  been observed experimentally under different physical circumstances \citep{Berg69,Grimes79,Winter91,Tan95,Drewsen98,Robertson99,Molhave00,Deshpande08,Staanum10}.  Moreover, an ensemble of strongly correlated micron-sized negative dust particles form dust Coulomb  crystals \citep{Chu94,Thomas94,Hayashi94,Barkan95,Fortov97,Mohideen98} when they were confined by the sheath parabolic potential in low-temperature laboratory dusty plasma discharges \citep{Shukla02,Shukla09}.
The condensation of charged dust particles occurs since the  dusty plasma $\Gamma_d$  (the ratio between the Coulomb energy between highly charged dust grains and the average dust particle kinetic energy) becomes relatively large due to the high dust charge state and low dust temperature. The attraction between like-charged dust particles forming dust Coulomb crystals may also be attributed to attractive forces arising from overlapping of  the dusty plasma Debye spheres \citep{Resendes}, ion focusing and wakefield effects \citep{Nambu,Vlad95,Shukla96}, and dust dipole-dipole  interactions \citep{Mohideen98,Shukla02,Shukla09}. The Cooper's pairing of charged dust particles, which are glued by ions,
led  to the discovery of a soft- condensed matter  of dust particle crystals in low-density and low-temperature classical plasma with Maxwell-Boltzmann distributions for electrons and ions. It turns out that several milestones were reached in the areas  of ordered crystalline structures composed of charged particles (e.g. an ensemble of in low-temperature physical systems electrons, ions, as well as charged colloidal and dust particles) in  physical systems, which share some common physics.

However, solid density plasmas are of fundamental importance for industrial applications (e.g. semiconductors, nano-diodes and metallic nanostructures for thin films), for inertial confinement fusion (ICF) schemes that utilizes high density compressed (HDC) plasmas, as well as
for planetary systems [e.g. the core of Jupiter \citep{Fortov09}] and superdense astrophysical objects (e.g the cores of white dwarf stars, warm dense matter). In dense plasmas, one has to account for degeneracy \citep{Chandrasekhar,Chandra} of electrons which obey the Fermi-Dirac distribution function. Correspondingly, quantum mechanical effects play a vital role since in such dense plasmas the Wigner-Seitz radius $d=(3/4\pi n_0)^{1/3}$ is comparable to the thermal de Broglie  wavelength $\lambda_B = \hbar/m V_T$ (which is a measure of the extent of electron wave functions), where $\hbar$ is Planck's constant divided by $2\pi$, $m$ the electron mass, $V_T =\sqrt{k_B T/m}$ the electron  thermal  speed due to random electron motions, and $k_B$ the Boltzmann constant. Also, in dense plasmas with degenerate electrons,  $\lambda_B$ turns out to be much smaller than the Landau length  $\lambda_L =e^2/k_B T$, which can be  conveniently expressed as $k_B T \ll e^2/a_B$, where $a_B =\hbar^2/me^2$ is the Bohr radius of a hydrogen atom. The electron degeneracy effects at nanoscales in dense plasmas can thus  be captured through the consideration of the Fermi-Dirac statistics for electrons with spin-$1/2$ (Fermions), and overlapping of electron wavefunctions due to Heisenberg's uncertainty principle and Pauli's exclusion principle, as well as electron-exchange and electron-correlations. Hence, there are quantum forces \citep{Chandrasekhar,Chandra,Wilhelm71,Gardner96,Manfredi01,Manfredi05,Shukla06,Shukla07,Shaikh07,NC08,Brodin08,Melrose08,Tsint09, Shukla10,Shukla11,Haas11,Vlad11,Tito11} associated with the quantum statistical electron pressure \citep{Chandrasekhar,Chandra,Landau80}, the quantum Bohm potential \citep{Wilhelm71,Gardner96,Manfredi01,Manfredi05,Shukla10,Shukla11,Haas11} through which degenerate electrons can tunnel through (often known as the quantum electron recoil effect), as well as the electron-exchange and electron-correlations potentials \citep{Brey,Hedin71,NC08}. It has been  found that the above mentioned  quantum forces acting on degenerate electrons in quantum plasmas introduce new  dispersive features to electron plasma oscillations \citep{Klimo52a,Klimo52b,Bohm52,Bohm53} with frequencies in the
x-ray regime, which can be assessed by using  collective x-ray spectroscopic scattering techniques \citep{Glen,Glen1}. In fact, \citet{Glen} have reported observations of electron plasma oscillations (EPOs) in warm dense matter (with the peak electron number  density  $\sim 3 \times 10^{23}$ cm$^{-3}$ and the equilibrium electron and ion temperatures of 12 eV ($\sim 1.5 \times 10^{5}$ degrees Kelvin). The latter is different from the electron Fermi temperature  $ T_F =(\hbar^2/2 k_B m_e)(3 \pi^2 n_0)^{2/3} \approx   1.7 \times 10^5$ degrees Kelvin, corresponding to an electron
number density $n_0\approx 2.5\times10^{23}\,\mathrm{cm}^{-3}$). Thus, the recent experiments of \citet{Glen} have unambiguously demonstrated the importance of  the quantum statistical pressure  and quantum electron recoil effects on the frequency spectra of EPOs \citep{Klimo52a,Klimo52b,Bohm52,Bohm53,Shukla10,Shukla11}, although a previous experimental investigation  \citet{Watanabe} had already established  the  quantum dispersion properties of EPOs in metals.

Very recently, \citet{Shukla12a,Shukla12b,Shukla12c} discovered an oscillating shielded Coulomb (OSC) potential, (also referred to as the Shukla-Eliasson attractive potential (SEAP) \citep{Akbari12}),  which is valid for the potential energy much  smaller than $k_B T$ and $mc^2$, around a stationary test ion  in an unmagnetized  quantum plasma, where $c$ is the speed of light in vacuum. The SEAP arises due to collective interactions between an ensemble  of degenerate electrons that shield an isolated ion at atomic dimensions. The profile of  the SEAP in  quantum plasmas resembles the Lennard-Jones (LJ) potential in atomic gases. The Shukla-Eliasson attractive force (SEAF), defined as minus the gradient of the SEAP, brings ions closer to form  ion clusters in  quantum plasmas. In this paper, we demonstrate the formation of ion clusters at atomic dimensions by performing computer simulations of the equations of motion for an ensemble of ions that are interacting with each other  through the SEAF.

\section{The SEAF}

\begin{figure}
\centering
\includegraphics[width=7cm]{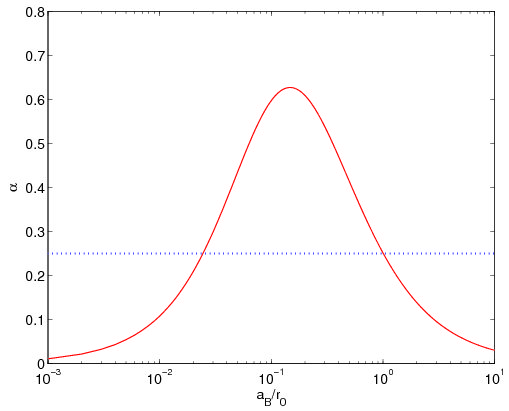}
\caption{The value of $\alpha$ as a function of $a_B/r_0$. The critical value $\alpha=1/4$ is indicated with a dotted line.}
\end{figure}

The existence of the SEAP, which is obtained from Fourier transformation of Poisson's equation
with the quasi-stationary electron density perturbation deduced from the linearized continuity and generalized momentum
equation \citep{Shukla12a,Shukla12b,Shukla12c} for non-relativistic, degenerate electrons in a dense quantum plasma,
critically depends on the electron number density through  the parameter $\alpha=\hbar^2 \omega_{pe}^2/4 m^2u_*^4$, where
$u_* =  (v_{*}^2/3+v_{ex}^2)^{1/2}$. The parameter $\alpha$ measures the quantum  electron recoil effect caused by the quantum Bohm potential \citep{Wilhelm71,Gardner96,Manfredi01,Manfredi05} $V_B = (\hbar^2/2m)(1/\sqrt{n}) \nabla^2 \sqrt{n}$ compared to the quantum statistical Fermi electron pressure  and the electron-exchange and electron-correlation effects arising from $1/2$-spin of the degenerate electrons.  Here  $v_*=\hbar (3\pi^2)^{1/3}/m  r_0$ is the electron Fermi  speed and   $v_{ex}= (0.328e^2/m r_0)^{1/2}[1+0.62/(1+18.36a_B n_0^{1/3})]^{1/2}$ includes the effects of electron exchange and electron correlations,  where $r_0 =n_0^{-1/3}$ represents the average inter-electron distance.
The expression  for $v_{ex}$ is derived by linearizing the sum of the electron exchange and  electron correlation potentials \citep{Brey,Hedin71}   $V_{xc} =0.985 e^2 n^{1/3} \left[1+ (0.034 /a_B n^{1/3}) {\rm ln}(1+18.37 a_B n^{1/3})\right]$. We note that $\alpha$ depends only on $a_B/r_0$, as  $\alpha\simeq 9.3\pi (a_B/r_0)/[ 1 + (3\pi^2)^{2/3} (a_B/r_0 )+0.62/(1+18.36 a_B/r_0)]^2$. \citet{Shukla12a,Shukla12b,Shukla12c} found that attractive potentials between ions exist only for $\alpha>1/4$. Figure 1 displays the value of $\alpha$ as a function of  $a_B/r_0$, where one observes that it is above  a critical value $0.25$ only for a limited range $2\times 10^{-2}< a_B/r_0 <1$,  corresponding to an electron number density in  the range $5.4\times 10^{19}\,\mathrm{cm}^{-3} < n_0 <   6.7 \times10^{24}\,\mathrm{cm}^{-3}$  (with $a_B=5.3\times10^{-9}\,\mathrm{cm}$). The maximum value is   $\alpha\approx 0.627$ at $a_B/r_0\approx 0.15$, corresponding  to the electron number density $n_0\approx 2\times 10^{22}\,\mathrm{cm}^{-3}$, a few times below solid densities. The validity of  the SEAP has been further  expanded \citep{Akbari12}  for wider density ranges by including Chandrasekhar's (\citeyear{Chandrasekhar,Chandra}) generalized pressure law for degenerate electron fluids and Salpeter's (\citeyear{Salpeter61}) electron-exchange  and electron-correlation potentials that are of astrophysical interest (e.g. the cores of white dwarf stars). The nonlinear shielding effects on the SEAF is discussed by \citet{PRE_Comment}.

\begin{figure}[htb]
\centering
\includegraphics[width=7cm]{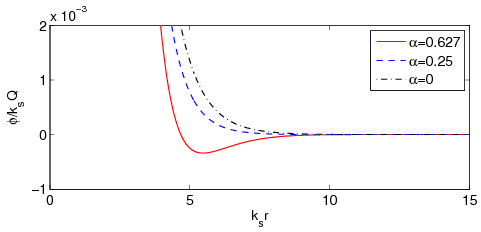}
\caption{(Color online) The electric potential $\phi$ as a function of $r$ for $\alpha=0.627$ (solid curve),
$\alpha=0.25$ (dashed curve) and $\alpha=0$ (dash-dotted curve). The value 0.627 is the maximum possible value of
$\alpha$ in our model, obtained for $a_B/r_0\approx 0.15$.}
\end{figure}

\begin{figure}[htb]
\centering
\includegraphics[height=7cm]{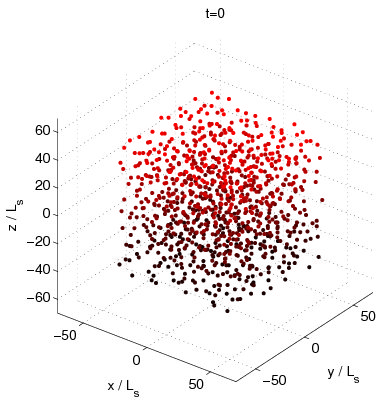}
\caption{(Color online) The initial positions of ions in the particle simulations.}
\end{figure}
\begin{figure*}[htb]
\centering
\includegraphics[height=7cm]{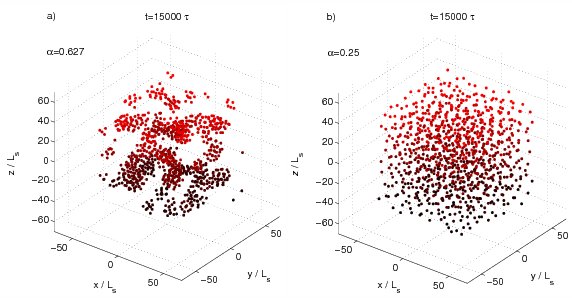}
\caption{(Color online) The positions of ions $t=15000$ for a) $\alpha=0.627$ and b) $\alpha=0.25$, showing the clustering and solidification
of  ions for $\alpha=0.627$, but not for $\alpha=0.25$.}
\end{figure*}

For $\alpha> 0.25$, the profile of the electric potential as a function of distance $r$ around a stationary test ion
charge $Q$ is \citep{Shukla12a,Shukla12b,Shukla12c}
%1
\begin{equation}
\phi({\bf r})= \frac{Q}{r} \big[ \cos(k_i r) + b_\ast \sin(k_i r) \big] \exp(-k_r r),
\label{pot1}
\end{equation}
which is referred to as the SEAP. Here $Q$ is the ion  charge, $b_\ast=1/\sqrt{4\alpha-1}$, $k_i=({k_s}/\sqrt{4\alpha})(\sqrt{4\alpha}-1)^{1/2}$, and $k_r=({k_s}/\sqrt{4\alpha})(\sqrt{4\alpha}+1)^{1/2}$,  with $k_s=\omega_{pe}/u_*$ being the modified inverse TF screening length.
The spatial profile of the SEAP in Fig. 2 shows a distinct minimum for the case $\alpha= 0.627$.
The negative part of the SEAP, given by  Eq. (1),  resembles the LJ potential and leads to
a short-range SEAF between neighboring ions.
On the other hand, the potential distribution around  a test  ion for $\alpha < 0.25$  reads \citep{Shukla12a,Shukla12b,Shukla12c}
%2
\begin{equation}
\phi({\bf r})=\frac{Q}{2 r} [(1+b)\exp(-k_{+}r) + (1-b) \exp(- k_{-} r)],
\end{equation}
where $b =1/\sqrt{1-4\alpha}$, and $k_\pm=k_s (1\mp\sqrt{1-4\alpha})^{1/2}/\sqrt{2\alpha}$. For this case, the potential
is positive and monotonically decreasing (cf. Fig. 2), giving rise to only a repulsive force (similar to the TF force) between ions.
In the high- or low-density limit,  where $\alpha \rightarrow 0$, we recover the  modified TF screened Coulomb
potential  $\phi({\bf r})= (Q/r) \exp(-k_s r)$.

\section{Demonstration of  ion clustering}

\begin{figure*}[htb]
\centering
\includegraphics[height=7cm]{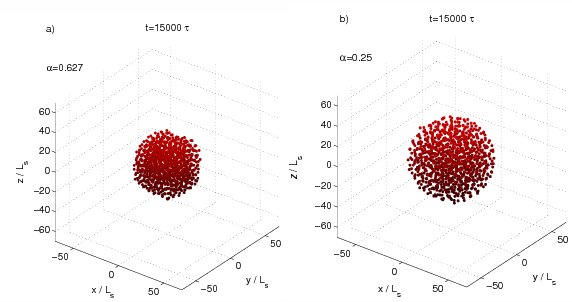}
\caption{(Color online) The positions of ions at $t=15000$ for a) $\alpha=0.627$ and b) $\alpha=0.25$ including a symmetric parabolic potential with $\omega_\perp=\omega_z=2\times10^{-3}\,\tau^{-1}$. Almost spherical non-Coulombic ion crystals are formed.}
\end{figure*}

We here present a computer simulation study of the dynamics of a system of ions interacting  with each other under the action of the SEAF.
For this purpose, we numerically solve the equations of motion for a system of ions with equal charges and masses, given by
%3
\begin{equation}
M \frac{d{\bf v}_j}{d t} = - Q \sum_{i\neq j} \nabla_j \phi(|{\bf R}_{ij}|)
-\nabla_jV_c({\bf r}_j)-M \nu {\bf v}_j,
\label{newton}
\end{equation}
where instantaneous position of each ion is determined from $d {\bf r}_j/dt={\bf v}_j$.
Here ${\bf R}_{ij} ={\bf r}_i - {\bf r}_j$ is the radius vector between particle $i$ and $j$,
${\bf r}_j(t)$ the position and ${\bf v}_j(t)$ the velocity of the $j$th ion, $M$ the ion mass,
$\nabla_j$ denotes the gradient of $\phi$ at position ${\bf r}_j$, and $\nu$ denotes an effective
collision frequency, which tends to retard the ion motion. The external confining potential,
$V_c({\bf r}) = (M/2) (\omega_\perp^2 r_\perp^2 + \omega_z^2 z^2$),  of charged particles
may have different  amplitudes $\omega_\perp$  and $\omega_z$ perpendicular and parallel
to the $z$-axis, respectively,  where $r_\perp^2=r_x^2+r_y^2$.

In order to demonstrate the clustering of ions under the SEAF, we now carry out particle simulations of  Eq. (3) with 1000 particles, initially  randomly placed in space, as shown in Fig. 3. In the first set of simulations, displayed in  Fig. 4, we consider interactions of the ions in the absence of the external confining potential $V_c$, viz. $\omega_z=\omega_\perp=0$,  and with $\nu=0.01\,\mathrm{\tau}^{-1}$. The positions of ions
at the end of the simulations are shown in Fig. 4 at time $t=15000\,\tau$. Here the positions and time are in units of $L_s=k_s^{-1}$ and  $\tau=M^{1/2}/Q k_s^{3/2}= (\pi/16)^{1/8}(a_B/r_0)^{3/8}\omega_{pi}^{-1}/\alpha^{3/8} Z_i$, respectively, where $\omega_{pi}=(4\pi Q^2 n_{i0}/M)^{1/2}$  is the ion plasma frequency, and $n_{i0}$  the equilibrium ion number density, related to the electron number density $n_0$ via the quasi-neutrality condition $Z_i n_{i0}=n_0$, where $Z_i$ is the ion charge state.  For $\alpha=0.627$,  which leads to a potential minimum
(cf. Fig. 2), we observe the clustering  of ions and the  formation of large-scale ionic structures. The clustering of ions is a relatively slow process in comparison with the ion plasma period $2\pi/\omega_{pi}$. Ion pairs and smaller clusters are initially formed, and later the larger ion clusters  are gradually formed by the agglomeration of smaller ion clusters. As a contrast, for $\alpha=0.25$, we see in Fig. 4 that there does not exist
condensation/coalescence of ions. This is due to the fact that there is no potential minimum for this value of  $\alpha$  (cf. Fig. 2),  and hence
the inter-ion force is always repulsive for this case.

\begin{figure*}[htb]
\centering
\includegraphics[height=7cm]{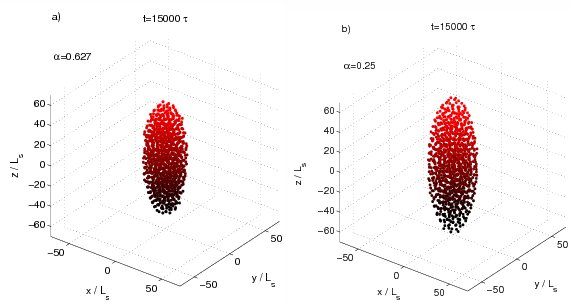}
\caption{(Color online) The positions of ions  at $t=15000$ for a) $\alpha=0.627$ and b) $\alpha=0.25$ including an asymmetric parabolic potential with $\omega_\perp=6\times 10^{-3}\,\tau^{-1}$ and $\omega_z=2\times10^{-3}\,\tau^{-1}$.  Here elongated non-Coulombic ion crystals
are formed.}
\end{figure*}

\begin{figure*}[htb]
\centering
\includegraphics[height=7cm]{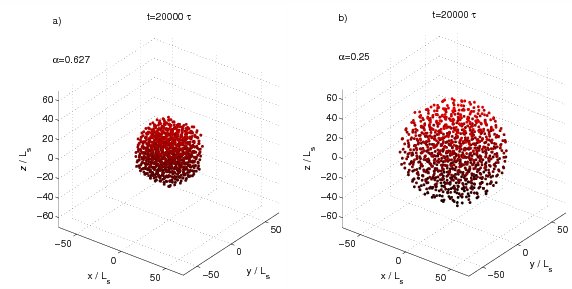}
\caption{(Color online) The positions of ions at $t=20000\tau$ for a) $\alpha=0.627$ and b) $\alpha=0.25$
for the case of a symmetric parabolic potential in Fig. 5, where the potential is set to zero at $t=15000\tau$.
The ion cluster remains tightly packed for $\alpha=0.627$, while the ion cloud expands for $\alpha=0.25$.}
\end{figure*}

\begin{figure*}[htb]
\centering
\includegraphics[height=7cm]{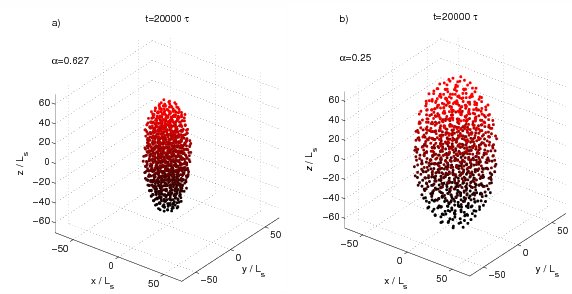}
\caption{(Color online) The positions of ions at $t=20000\tau$ for a) $\alpha=0.627$ and b) $\alpha=0.25$
for the case of an asymmetric parabolic potential in Fig. 6, where the potential is set to zero at $t=15000\tau$.
The ion cluster remains tightly packed for $\alpha=0.627$, while the ion cloud expands for $\alpha=0.25$.}
\end{figure*}

Furthermore, we  have carried  out a set of simulations with a symmetric potential $V_c$ with $\omega_\perp=\omega_z=2\times 10^{-3}\,\tau^{-1}$.  Here, as seen in Fig. 5, almost spherical  non-Coulombic ion crystals are formed for both $\alpha=0.627$ and $\alpha=0.25$.
We also performed simulations including an asymmetric external potential $V_c$ with $\omega_\perp=6\times 10^{-3}\,\tau^{-1}$
and  $\omega_z=2\times 10^{-3}\,\tau^{-1}$, providing a stronger confinement in the perpendicular direction. Figure 6 displays the final state
 for $\alpha=0.627$ and $\alpha=0.25$, where ions, in both cases,  form non-Coulombic ion crystals elongated along the $z$-direction.
In general,  the formation of non-Coulombic ion crystals is attributed  to the balance between the external and inter-ion potentials, where the system  tends to a configuration of a minimum potential energy. We note that similar configurations have been previously reported  for both charged  macro-particles \citep{Wuerker59} and ions \citep{Drewsen98,Molhave00,Staanum10} confined by external potentials in the Paul trap.
Finally, in Figs. 7 and 8, we continued the simulations in Figs. 5 and 6 and set the external confining potential to zero at time $t=15000\tau$.
For both cases, with symmetric and asymmetric potentials, we  see that ion crystals remained tightly packed for $\alpha=0.627$, where it performed damped oscillations as elastic solids before settling down to the final states in Figs. 7(a) and 8(a). In contrast, for $\alpha=0.25$,
the ion cloud expanded to larger and less dense ion clouds, as seen in Figs 7(b) and 8(b) at $t=20000\tau$. Hence the short-range potential
around ions for $\alpha=0.627$ (cf. Fig. 2) lead to the solidification of the ions, which is not the case for the the short-range repulsive potential
for $\alpha=0.25$.

\section{Summary and conclusions}

In this paper, we have carried out particle simulations to demonstrate clustering of ions due to the  newly found SEAF arising from collective interactions between an ensemble of degenerate electrons that shield ions in dense quantum plasmas. Specifically, the SEAF leads to clustering/condensation or coagulation of ions in the absence of an external confining potential for charged particles. However, ion clustering can be put on the firm footing by calculating the dynamical ion structure factor (DISF) based on the
fluctuation-dissipation theorem and the dielectric constant of degenerate electrons and strongly correlated ions in a viscous quantum plasma. It may well turn out that the DISF will reveal long-range correlations between ions. It is our belief that the formation of ion nano-clusters is going to play a valuable role  in the area  of  high-density compressed plasmas with degenerate electrons \citep{Glen1,Son04,Son05,Son06,Malkin07} for ICF  to succeed, and also  in the  emerging field  of nano-material sciences (e.g.  nanodiodes, metallic nanostructures for thin films \citep{NC08}, nanowires), where  closely-packed ions will lend support  to enhanced fusion probabilities
(with anomalous fusion  cross-sections) for controlled thermonuclear ICF,   and may also influence the electric properties (e.g. resistivity)  of
new high density plasma materials at relatively high temperatures. Finally,  we stress that  the Cooper  pairing of ions at atomic dimensions shall provide possibility  of novel superconducting plasma-based nanotechnology, since  the electron transport in nanostructures would be  rapid due to shortened  distances between ions in the presence of the novel  SEAF.

\acknowledgments
This work was supported by the Deutsche Forschungsgemeinschaft through the project SH21/3-2 of the Research Unit 1048.

\end{document}